\documentclass[prd,aps,preprintnumbers, showpacs, nofootinbib,superscriptaddress,notitlepage]{revtex4-1}
 \usepackage{latexsym,bm,amsmath,amssymb,amsfonts}
\newcommand{\ave}[1]{\left\langle #1 \right\rangle}
\newcommand{\order}[1]{ \mathcal{O} \left( #1 \right) }

  \usepackage{latexsym,bm,amsmath,amssymb,amsfonts}
  \usepackage{epsfig,graphics,graphicx}
  \usepackage{slashed}
\usepackage{latexsym,bm,amsmath,amssymb,amsfonts}

\newcommand{\nn}{\nonumber \\}
\newcommand{\beq}{\begin{eqnarray}}
\newcommand{\eeq}{\end{eqnarray}}

\begin{document}

\preprint{YITP-14-59}

\title{Flow harmonics within an analytically solvable viscous hydrodynamic model}
\author{Yoshitaka Hatta}
\affiliation{Yukawa Institute for Theoretical Physics, Kyoto University, Kyoto 606-8502, Japan}

\author{Jorge Noronha}
\affiliation{Instituto de F\'isica, Universidade de S\~ao Paulo, C.P. 66318, 05315-970 S\~ao Paulo, SP, Brazil}

\author{Giorgio Torrieri}
\affiliation{IFGW, Universidade Estadual de Campinas, Campinas, S\~ao Paulo, Brazil}
\author{Bo-Wen Xiao}
\affiliation{Key Laboratory of Quark and Lepton Physics (MOE) and Institute
of Particle Physics, Central China Normal University, Wuhan 430079, China}

\date{\today}
\vspace{0.5in}
\begin{abstract}
Based on a viscous hydrodynamic model with anisotropically perturbed Gubser flow and isothermal Cooper-Frye freezeout at early times, we analytically compute
the flow harmonics $v_n(p_T)$ and study how they scale with the harmonic number $n$ and transverse momentum, as well as
the system size, shear and bulk viscosity coefficients, and collision energy. In particular, we find that the magnitude of shear viscous corrections grows linearly with $n$. The mixing between different harmonics is also discussed. While this model is rather simple as compared to realistic heavy-ion collisions, we argue that the scaling results presented here may be meaningfully compared to experimental data collected over many energies, system sizes, and geometries.
\end{abstract}
\maketitle

\section{Introduction}
The hydrodynamic model is reckoned to be phenomenally successful at describing a lot of data measured in heavy-ion collisions 
\cite{sqgpmiklos,sqgpshuryak,hwa,whitebrahms,whitephobos,whitestar,whitephenix,alice,cms,atlas}.
The main phenomenological application of hydrodynamics in heavy-ion physics is, given initial conditions, an equation of state and transport coefficients, to calculate the azimuthal structure of transverse momentum of particles, usually parametrized by Fourier coefficients $v_n$
\begin{equation}
\label{v2def}
E_p\frac{dN}{d^3 p} = \frac{dN}{2\pi p_T dp_T dY} \left\{1+2 \sum_n v_n(p_T,Y) \cos \left[n (\phi_p-\Psi_n)\right] \right\}\,,
\end{equation}
where $\Psi_n$ are reaction planes characterizing the event \cite{Poskanzer:1998yz}, $\phi_p$ is the azimuthal angle, and $Y$ is the momentum rapidity of the observed particles.

At RHIC and LHC, with reasonable initial conditions, an equation of state
 compatible with lattice QCD and the viscosity to entropy density ratio $\eta/s \sim 0.2$, $v_2$ is reproduced quantitatively and $v_{n=3-5}$ qualitatively agrees (see the recent review in \cite{Heinz:2013th}).  While the full hydrodynamic model has a number of a priori undetermined parameters 
 the abundance and quality of experimental data in azimuthal particle distributions make the determination of most relevant parameters possible, 
in analogy to how the cosmic microwave background perturbations can constrain inflaton potentials in the big bang theory.

Unlike the big bang which is unique, 
the little bang \cite{littlebang, Heinz:2013wva} occurred in ultrarelativistic heavy-ion collisions can be repeated many times, 
and the events can come in many different shapes and sizes.
Experimentalists have been able to vary collision energy, system size, rapidity and initial geometry, and to study the system response for all these variables. This combined analysis has revealed that $v_n(p_T)$ exhibits a remarkably simple scaling behavior \cite{whitephobos,gtscaling} in a way that might require some intuitive understanding rather than sophisticated fitting.

 Such intuitive understanding
has been historically closely associated with fluid mechanics \cite{pihydro}. The well-known Bjorken flow solution \cite{bjorken} indicates one type of scaling which relates the initial temperature $T_i$ and the start of the hydrodynamic evolution $\tau_i$ to the final multiplicity $dN/dY$ and transverse area $A_\perp$
\begin{equation}
\label{bjorken}
\frac{dN}{dY} \sim \tau_i  T_i^3 A_\perp \,.
\end{equation}
This formula has been used by experimentalists to make an order of magnitude estimate of the initial temperature for given colliding systems. 
However, since the dynamics is purely longitudinal, it does not provide us with the necessary transverse information which is essential for analyzing flow harmonics $v_n$. 

Recent progress in applying conformal geometry to hydrodynamic analytical solutions \cite{Gubser:2010ze,Gubser:2010ui} could however yield ``toy models'' which are sophisticated enough to incorporate realistic physics and yet being analytically solvable. This work goes in this direction, by applying these analytical solutions to ``realistic'' geometries and freezeout criteria which makes experimental comparisons possible. 
(See \cite{Csanad:2005gv} for an approach similar in spirit using a non-boost-invariant hydrodynamic solution.)

The model presented in this paper incorporates longitudinal flow, transverse flow, and flow anisotropies. As a result, it is sensitive not only to longitudinal parameters in a similar way as the Bjorken flow solution \cite{bjorken}, but also to transverse geometry and first order (Navier-Stokes) transport coefficients.
The main objective  is to investigate, within this model,
 the analytical behavior of flow harmonics as functions of harmonic number $n$, transverse momentum $p_T$ as well as shear and bulk viscosity coefficients.
    The assumptions involved are: (i) conformal symmetry which QCD approximately possesses  at high temperature; (ii) an early freezeout along an isothermal hypersurface; (iii) the Cooper-Frye formula for the particle distribution \cite{cooperfrye} augmented with viscous corrections to the distribution function \cite{teaney}. Under these assumptions, we shall calculate $v_n(p_T)$ for central and semi-central collisions where $v_n$ is linearly proportional to the eccentricity.  We hope to be able, in this way, to study the
 response of the system to initial geometry changes and the emerging scaling behavior
  in terms of the bulk characteristics of the event.

This paper is organized as follows. In the next section we define the anisotropic perturbation around Gubser flow used in this paper. In Section III, we present the necessary details involving the Cooper-Frye freeze-out (including shear viscosity effects) that are needed to compute the anisotropic flow coefficients, as shown in Sec.\ IV. In the following section we discuss the mixing between different harmonics $v_n$ and $v_{2n}$ (for an inviscid fluid) in our approach. Section VI is devoted to the study of the effects of bulk viscosity, treated as a small perturbation in an ideal fluid, on the flow harmonics. In Sec.\ VII, we discuss how the analytical expressions for the anisotropic flow coefficients found in the previous sections can be useful to study how these quantities may scale with experimental observables such as the particle multiplicity, collision energy, etc. We finish with our conclusions and outlook in Sec.\ VIII. We use the mostly plus metric signature which means that the flow velocity in hydrodynamics obeys $u_\mu u^\mu=-1$.

\section{Anisotropic perturbation around Gubser flow}

In Ref.~\cite{Gubser:2010ze}, Gubser obtained an exact boost-invariant solution of the relativistic conformal Navier-Stokes (NS) equation that has a nontrivial transverse profile. Subsequently, Gubser and Yarom \cite{Gubser:2010ui} considered anisotropic perturbations on top of the solution. In this section we first show that, in the early time regime, these (boost invariant) perturbed solutions can be described fully analytically.

Consider a conformal theory in which the energy density $E$, the pressure $P$ and the temperature $T$ are related as $E=3P=\lambda T^4$ ($\lambda$ is a constant which depends on the theory of interest).
We work in the coordinate system
\beq
ds^2 = -d\tau^2 + dx_\perp^2 + x_\perp^2 d\phi^2 + \tau^2 dy^2\,,
\eeq
where $\tau$ is the proper time and $y$ is the spacetime rapidity.
The unperturbed solution of the NS equation for the energy density reads \cite{Gubser:2010ze,Gubser:2010ui}
\beq
E_{NS}=\lambda T_{NS}^4=\frac{1}{\tau^4}\frac{\lambda C^4}{(\cosh\rho)^{8/3}}\left[1+\frac{\eta_0}{9\lambda C}(\sinh\rho)^3 \,_2F_1\left(\frac{3}{2},\frac{7}{6},\frac{5}{2};-\sinh^2\rho\right)\right]^4\,, \label{gub}
 \eeq
  where $C$ is a dimensionless normalization factor, $\,_2F_1$ is the hypergeometric function and $\rho$ is defined by
\beq
 \sinh\rho \equiv-\frac{L^2- \tau^2+x_\perp^2}{2L\tau}\,. \label{L}
 \eeq
In (\ref{L}),  $L$ is a length parameter which roughly characterizes the initial transverse size of the flow, and $\eta_0 \equiv \eta/T^3$ is the rescaled,  dimensionless shear viscosity. The flow velocity $u^\mu$ is given by
\beq
u^{\tau}_{NS}=\cosh\left[\tanh^{-1}\frac{2\tau x_\perp}{L^2+\tau^2
+x_\perp^2}\right]\,,\quad
u^{\perp}_{NS}=\sinh\left[\tanh^{-1} \frac{2\tau x_\perp}{L^2+\tau^2+x_\perp^2}\right]\,, \label{boo}
\eeq
 with $u^\phi_{NS}=u^y_{NS}=0$. We shall focus on the early time regime  $\tau \ll L$ in which  (\ref{boo}) reduces to
 \beq
 u^{\tau}_{NS} \approx 1+{\mathcal O}(\tau^2)\,, \qquad u^{\perp}_{NS} \approx \frac{2\tau x_\perp}{L^2+x_\perp^2}\,.
 \eeq
  In this regime, the expansion of the flow is predominantly longitudinal, while the transverse velocity is small and not yet fully developed $|u^\perp|\ll 1$.
We neglect ${\mathcal O}(\tau^2)$ corrections to $u^\tau$ so that $u^\tau \approx 1$ throughout this paper. Clearly, these approximations break down when $\tau \sim {\mathcal O}(L)$.

Following \cite{Gubser:2010ui}, we consider  rapidity-independent perturbations of the form
\beq
T&=&T_{NS}(1+ {\mathcal S}\delta)\,, \\
u_\perp &=& u^{\perp}_{NS} + \frac{2\tau L}{L^2+x_\perp^2} \nu_s \partial_\Theta {\mathcal S}\,,\\
u_\phi &=&\tau \nu_s \partial_\phi {\mathcal S}\,, \label{perturb}
\eeq
 where $\delta$ and $\nu_s$ are the fluctuations of the temperature and the velocity, respectively.
  ${\mathcal S}$ is a linear combination of the spherical harmonics $Y_{lm}(\Theta,\phi)$ where the polar coordinates $(\Theta,\phi)$ refer not to the ordinary three dimensional space, but to the de Sitter space to which the Minkowski space is conformally related. We take
\beq
{\mathcal S} =-\epsilon_n
\left( \frac{2Lx_\perp}{L^2+x_\perp^2}\right)^n \cos n\phi \propto Y_{n,n}(\Theta,\phi)+Y_{n,-n}(\Theta,\phi)\,, \label{y}
\eeq
 which gives rise to the flow harmonics $v_n$. $\epsilon_n$ is the corresponding eccentricity (see below).
 The equations for  $\delta$ and $\nu_s$  are obtained by linearizing the Navier-Stokes equation and are given in  \cite{Gubser:2010ui}. While these equations are difficult to handle analytically in the viscous case, we point out that, in the early time regime $\tau \ll L$ (or $\rho \to -\infty$), and to linear order in $\eta_0$, they take the following simple form
 \beq
\frac{d}{d\rho} \begin{pmatrix} \delta \\ \nu_s \end{pmatrix}=-\begin{pmatrix}
\frac{\eta_0}{3\lambda C}\left(2e^\rho\right)^{-2/3} & {\mathcal O}(e^{2\rho}) \\
1+\frac{\eta_0}{\lambda C} (2e^{\rho})^{-2/3} & \frac{2}{3}+\frac{\eta_0}{\lambda C} (2e^\rho)^{-2/3}
\end{pmatrix} \begin{pmatrix} \delta \\ \nu_s \end{pmatrix}\,.
\label{l}
\eeq
The approximate solution is
\beq
\delta &=& 1 +\frac{\eta_0}{2\lambda C}(2e^\rho)^{-2/3} + {\mathcal O}(e^{2\rho})\approx 1 +\frac{\eta_0}{2\lambda C}\left(\frac{L^2+x_\perp^2}{2L\tau}\right)^{2/3}\,,\\
\nu_s &=& -\frac{3}{2}+ k\frac{\eta_0}{\lambda C} e^{-2\rho/3}+{\mathcal O}(e^{2\rho}) \,,
\eeq
 where $k$ is an arbitrary constant. We choose $k=0$ for simplicity.
 We thus arrive at the following perturbed solution
\beq
E
&\approx &\frac{\lambda C^4}{\tau^{4/3}}\frac{(2L)^{8/3}}{(L^2+x_\perp^2)^{8/3}} \left(1-\frac{\eta_0}{2\lambda C}\left(\frac{L^2+x_\perp^2}{2L\tau}\right)^{2/3} \right)^4
\nn && \qquad \quad  \times \left[1-4\epsilon_n \left(1+\frac{\eta_0}{2\lambda C}\left(\frac{L^2+x_\perp^2}{2L\tau}\right)^{2/3} \right) \left( \frac{2Lx_\perp}{L^2+x_\perp^2}\right)^n \cos n\phi \right]\,,
\nn
u_\perp &=& \frac{2\tau x_\perp}{L^2+x_\perp^2} + \epsilon_n \frac{3nL\tau }{L^2+x_\perp^2} \left( \frac{2Lx_\perp}{L^2+x_\perp^2}\right)^{n-1}
\frac{L^2-x_\perp^2}{L^2+x_\perp^2} \cos n\phi\,,\nn
u_\phi &=& -\epsilon_n\frac{3n\tau }{2}  \left( \frac{2Lx_\perp}{L^2+x_\perp^2}\right)^n  \sin n\phi\,. \label{new}
\eeq
When $n=2$ (elliptic flow) and in the large-$x_\perp$ region $x_\perp \gg L$, (\ref{new}) coincides with the approximate solution constructed in \cite{Hatta:2014upa} via a different method.  However, the present solution is better-behaved at small-$x_\perp$. Note that the coefficient of $\cos n\phi$ in $u_\perp$ becomes negative at large distances $x_\perp >L$, which is  characteristic of conformal solutions  \cite{Hatta:2014upa}.

The parameter $\epsilon_n$ can be identified with the eccentricity which we define as
\beq
\epsilon_n \propto -\frac{\int d^2x_\perp E^{3/4}\frac{x_\perp^n}{(L^2+x^2_\perp)^{n-1}} \cos n\phi }{\int d^2x_\perp E^{3/4} \frac{x_\perp^n}{(L^2+x_\perp^2)^{n-1}} }\,, \label{ex}
\eeq
 where the weight $E^{3/4}$ has conformal dimension three and is essentially the entropy density. (\ref{ex}) differs from the more common definition of $\epsilon_n$ by a factor $(L^2+x_\perp^2)^{n-1}$ in the integrand. Without this factor, the denominator of (\ref{ex}) is divergent because $E$ has a power-law fall-off in a conformal theory and one has to introduce a cutoff. Our choice is motivated by the discussion in Ref.~\cite{Gubser:2010ui} about the preferred way of defining various anisotropic moments in a conformal theory. One can check that the right-hand-side of (\ref{ex}) computed with the energy density in (\ref{new}) is $\epsilon_n$ times a function which is  approximately independent of $n$.\footnote{There are other cutoff-independent definitions of $\epsilon_n$. For example, one may try
  \beq
 -\frac{\int d^2x_\perp Ex_\perp^2 \cos n\phi }{\int d^2x_\perp E x_\perp^2} \,.
\eeq
 With $E$ as given by (\ref{new}), this integral behaves like $\epsilon_n/\sqrt{n}$, so  one has to redefine $\epsilon_n \to \sqrt{n}\epsilon_n$ in (\ref{y}). This  shows that the power of $n$ in the prefactor of the formulas below depends on how we define $\epsilon_n$.
}

\section{Cooper-Frye formula}

  We use the solution (\ref{new}) as a way of evaluating the flow harmonics $v_n(p_T)$. Of course, $v_n$ is a feature of the final state of heavy-ion collisions, whereas (\ref{new}) is valid only at early times $\tau \ll L$ where $L$ may roughly be thought of  as the nuclear diameter. Nevertheless, we define a `time-dependent' $v_n(p_T,\tau)$ via  the Cooper-Frye formula \cite{cooperfrye} assuming that the system freezes out at any instant of time including very early times. Specifically, we shall compute
\beq
(2\pi)^3 \frac{dN}{dY p_T dp_T d\phi_p} = - \int_\Sigma p^\mu d\sigma_\mu f(p^\mu u_\mu/T)  \  \propto 1+2v_n(p_T,\tau) \cos n\phi_p\,, \label{coop}
\eeq
  where $f$ is the distribution function and the integral is taken over the isothermal surface $\Sigma$. From here on, $T$ denotes the freezeout temperature.

First let us consider the flow profile at the freezeout. From our solution, it is easy to analytically determine the isothermal surface $T(x_\perp,\phi)=const.$
\beq
T^3&=& \frac{C^3(2L)^{2}}{\tau (L^2+x_\perp^2)^2}\left(1-\frac{\eta_0}{2\lambda C} \left(\frac{L^2+x_\perp^2}{2L\tau}\right)^{2/3}\right)^3 \nn
&& \times \left[1-3\epsilon_n \left(1+\frac{\eta_0}{2\lambda C}\left(\frac{L^2+x_\perp^2}{2L\tau}\right)^{2/3} \right) \left( \frac{2Lx_\perp}{L^2+x_\perp^2}\right)^n \cos n\phi \right] \equiv \frac{C^3B^3}{(2L)^3}\,,
\label{free}
\eeq
 where $B$ is a dimensionless parameter.
To linear order in $\eta_0$ and $\epsilon_n$, (\ref{free}) can be iteratively solved for the freezeout time
\beq
\tau(x_\perp,\phi) &\approx& \frac{(2L)^5}{B^3(L^2+x_\perp^2)^2}\left(1-\frac{3\kappa  (L^2+x_\perp^2)^2}{2(2L)^4} -3\epsilon_n \left( \frac{2Lx_\perp}{L^2+x_\perp^2}\right)^n    \cos n\phi   \right)\,, \nn
 &\equiv& \tau_0+\delta \tau  \epsilon_n\cos n\phi\,, \label{time}
 \eeq
 where the coefficient of the ${\mathcal O}(\eta_0 \epsilon_n)$ term has been exactly canceled.
 In (\ref{time}), we defined\footnote{Usually the Knudsen number is defined as the ratio of microscopic to macroscopic quantities.   If $L$ were the only macroscopic quantity, this would imply that $\kappa \sim (TL)^{-1}(\eta/S)$, but our system also assumes $\tau \ll L$.  Since Eq.~(\ref{free}) implies $T \sim BC/L$ and Eq.~(\ref{time}) implies $\tau \sim L/B^3$, our definition is compatible with the standard one provided $\tau$ is the macroscopic quantity in the denominator.
   Note that $\sigma^\mu_{\ \nu}, \nabla_\mu u^\mu \sim {\mathcal O}(1/\tau)$ at early times, so it is indeed more appropriate to use $\tau$ as the relevant scale. } the `Knudsen number' \cite{Hatta:2014upa}
  \beq
  \kappa \equiv \frac{\eta_0 B^2}{\lambda C} = \frac{4B^2}{3C}\frac{\eta}{S}\,,
  \label{kn}
  \eeq
    where $S=(E+P)/T$ is the entropy.
  From (\ref{time}), we see that the condition  $\tau \ll L$ implies  $B^3\gg 1$, and the $\tau$-dependence of $v_n(p_T,\tau)$ is effectively converted to the $B$-dependence $v_n(p_T, B)$.
We denote the flow velocity {\it on the isothermal surface} as
 \beq
  u_\perp \equiv u_{\perp 0}+\delta u_\perp \epsilon_n \cos n\phi\,, \qquad u_\phi \equiv \delta u_\phi \epsilon_n \sin n\phi\,.
  \label{utau}
 \eeq
This can be obtained by substituting (\ref{time}) into (\ref{new}), and the result is
\beq
u_{\perp 0} &=& 2x_\perp \frac{(2L)^5}{B^3(L^2+x_\perp^2)^3}(1-\alpha)\,, \label{u0}\\
\delta u_{\perp } &=& \frac{3(2L)^5}{B^3(L^2+x_\perp^2)^4}\left( \frac{2Lx_\perp}{L^2+x_\perp^2}\right)^{n-1} L\left(n(L^2-x_\perp^2)(1-\alpha)-4x_\perp^2\right)\,,  \\
 \delta u_\phi &=& -\frac{3n}{2}\frac{(2L)^5}{B^3(L^2+x_\perp^2)^2}\left( \frac{2Lx_\perp}{L^2+x_\perp^2}\right)^{n}(1-\alpha)\,,
\eeq
 where
 \beq
 \alpha\equiv \frac{3\kappa (L^2+x_\perp^2)^2}{2(2L)^4}\,. \label{al}
 \eeq
 Also, the following combination
 \beq
\frac{1}{u_{\perp 0}}\left( \delta u_{\perp }-\frac{\delta u_\phi}{x_\perp} \right) = \frac{3}{ 1-\alpha}\left( \frac{2Lx_\perp}{L^2+x_\perp^2}\right)^{n}\left( \frac{nL^2(1-\alpha)}{2x_\perp^2} -1\right)\,
 \label{neg}
 \eeq
 will be relevant later in the next section.

Next we consider the distribution function $f$. In the presence of viscosity, $f$ has both the equilibrium and non-equilibrium parts $f=f_{eq} + \delta f$. We assume that the equilibrium part is the Boltzmann distribution and write
\beq
f =
e^{p^\mu u_{\mu}/T}
  \left(1+ \frac{p_\mu p_\nu \pi^{\mu\nu}}{2(E+P)T^2}\chi(p)\right)\,, \label{non}
\eeq
where
\beq
p^\mu u_\mu = -m_T \cosh (y-Y) + p_T u_\perp \cos(\phi-\phi_p) - \frac{p_T u_\phi}{x_\perp} \sin (\phi-\phi_p)\,.
\eeq
The transverse mass $m_T$ is defined as usual $m_T \equiv \sqrt{m^2+p_T^2}$ with $m$ being the particle mass. Strictly speaking, since we assume conformal symmetry, $m$ should be zero. However, for phenomenological purposes we take $m$ as a free parameter.
The function $\chi(p)$ is not unique and depends on the theory under consideration. For simplicity, we choose $\chi=1$, which is known as the democratic Ansatz \cite{Molnar:2011kx}, but other choices (see, \cite{Dusling:2009df,Bhalerao:2013pza}) can be dealt with.
 Then the second factor in (\ref{non}) becomes
\beq
&&\frac{p_\mu p_\nu \pi^{\mu\nu}}{2(E+P)T^2}  =-p_\mu p_\nu \sigma^{\mu\nu} \frac{3\eta_0 }{4\lambda T^3}
\approx \frac{\kappa (L^2+x_\perp^2)^2}{2^6T^2 L^4}\frac{\tau_0}{\tau} (p_T^2-2m_T^2 \sinh^2(y-Y))\,,
\label{bra}
\eeq

where we used $\pi^{\mu\nu}=-2\eta\sigma^{\mu\nu}$ and the fact that, at early  times, the dominant  components of the shear tensor $\sigma^{\mu\nu}$ are
\beq
\sigma^{\perp\perp}\approx x_\perp^2\sigma^{\phi\phi}\approx -\frac{1}{3\tau} + {\mathcal O}(\tau)\,, \qquad \tau^2\sigma^{\eta\eta}\approx \frac{2}{3\tau} + {\mathcal O}(\tau)\,,
\eeq
  with $\tau$ as given by (\ref{time}).

Finally, the integration measure in  (\ref{coop}) can be written as
\beq
-p^\mu d\sigma_\mu
 =x_\perp \tau \left(m_T \cosh(y-Y)  -p_T \cos (\phi-\phi_p) \frac{\partial \tau}{\partial x_\perp}+ \frac{p_T}{x_\perp}\sin (\phi-\phi_p) \frac{\partial \tau}{\partial \phi} \right)dy dx_\perp d\phi\,.
\label{coo}
\eeq

\section{Calculation of $v_n(p_T)$}

We now have all the necessary ingredients to evaluate the integral (\ref{coop}).
The $y$-integral can be easily done

\beq
&&(2\pi)^3\frac{dN}{dY p_T d p_T d\phi_p} = 2\int dx_\perp d\phi \, \exp\left(\frac{ p_T u_\perp}{T} \cos(\phi-\phi_p) - \frac{p_T u_\phi}{x_\perp T} \sin (\phi-\phi_p)  \right)  \nn
&& \qquad \times  x_\perp \tau\Biggl[m_T K_1( m_T/T) \left(1-\beta_\pi \frac{\tau_0}{\tau}\right) \nn
&& \qquad \qquad +K_0(m_T/T)\left(-p_T \cos (\phi-\phi_p) \frac{\partial \tau}{\partial x_\perp}
+ \frac{p_T}{x_\perp}\sin (\phi-\phi_p) \frac{\partial \tau}{\partial \phi}\right) \left(1-\widetilde{\beta}_\pi \frac{\tau_0}{\tau}\right)
\Biggr]\,, \label{eta}
\eeq
where we abbreviated
\beq
\beta_\pi \equiv \frac{\kappa (L^2+x_\perp^2)^2}{2^6 L^4}\left(\frac{m_T^2}{2T^2}\frac{K_3-K_1}{K_1}-\frac{p_T^2}{T^2}\right) \approx \frac{\kappa (L^2+x_\perp^2)^2}{2^6 L^4}\left(\frac{2m_T}{T}+3-\frac{p_T^2}{T^2}\right)\,, \nn
\widetilde{\beta}_\pi \equiv  \frac{\kappa (L^2+x_\perp^2)^2}{2^6 L^4}\left(\frac{m_T^2}{T^2}\frac{K_2 -K_0}{K_0}-\frac{p_T^2}{T^2}\right) \approx  \frac{\kappa (L^2+x_\perp^2)^2}{2^6 L^4}\left(\frac{2m_T}{T}+1-\frac{p_T^2}{T^2}\right)\,. \label{api}
\eeq
 The subscript $\pi$ used above is a reminder that these terms come from the $\delta f\sim \pi^{\mu\nu}$ term in (\ref{non}). The last expressions in (\ref{api}) are valid
 when $m_T\gg T$.

Next we expand the Boltzmann exponential  factor to linear order in $\epsilon_n$ using (\ref{utau}) and perform the $\phi$-integral. It is convenient to divide the integral into three parts corresponding to the three terms in (\ref{coo}).
\beq
(2\pi)^3\frac{dN}{dY p_T d p_T d\phi_p} \equiv J_1+J_2+J_3\,.
\eeq
The first integral reads
\beq
 J_1
 &=& 4\pi m_T  K_1(m_T/T)\int_0^\infty dx_\perp x_\perp \tau_0  \Biggl\{I_0(z) \left(1-\beta_\pi \right) +\frac{\delta \tau}{\tau_0}I_n(z)\epsilon_n\cos n\phi_p \nn
 && \qquad \qquad +(1-\beta_\pi) \frac{p_T}{2T}\left[ \left( \delta u_{\perp} -\frac{\delta u_\phi}{x_\perp}\right) I_{n-1}(z) +\left(\delta u_\perp +\frac{\delta u_\phi}{x_\perp}\right) I_{n+1}(z)\right] \epsilon_n \cos n\phi_p\Biggr\} \nn
 &\equiv &J_1^0+\delta J_1 \epsilon_n \cos n\phi_p\,,
  \label{cooper}
\eeq
  where we defined
  \beq
   z \equiv \frac{p_T u_{\perp 0}}{T} = \frac{2x_\perp p_T (2L)^5}{TB^3(L^2+x_\perp^2)^3}\left(1-\alpha \right)\,.
   \eeq
Similarly, the second integral is
\beq
J_2&= &-4\pi p_T K_0(m_T/T) \int_0^\infty dx_\perp x_\perp \tau_0  \Biggl\{\frac{\partial \tau_0}{\partial x_\perp} I_1(z) \left(1-\widetilde{\beta}_\pi \right) +\frac{\partial \tau_0}{\partial x_\perp} \frac{\delta \tau}{\tau_0}I'_n(z)  \epsilon_n \cos n\phi_p \nn
&& +(1-\widetilde{\beta}_\pi) \Biggl[\frac{\partial \tau_0}{\partial x_\perp} \frac{p_T}{2T}\left( \left( \delta u_{\perp} -\frac{\delta u_\phi}{x_\perp}\right) I'_{n-1}(z) +\left(\delta u_\perp +\frac{\delta u_\phi}{x_\perp}\right) I'_{n+1}(z)\right) + \frac{\partial \delta \tau}{\partial x_\perp} I'_n(z)  \Biggr]\epsilon_n \cos n\phi_p
 \Biggr\} \nn
&\equiv &J_2^0+\delta J_2 \epsilon_n \cos n\phi_p\,.
\eeq
The third integral is proportional to $\frac{\partial \tau}{\partial \phi} \sim \epsilon_n$, so one can set $\epsilon_n=0$ elsewhere
\beq
J_3&= &-4\pi p_T K_0(m_T/T)\int_0^\infty dx_\perp  \tau_0 \frac{n^2 \delta \tau}{z}I_n(z) (1-\widetilde{\beta}_\pi) \epsilon_n \cos n\phi_p \nn
&\equiv &\delta J_3 \epsilon_n \cos n\phi_p\,.
\eeq

From these integrals, the differential flow harmonics $v_n(p_T)$ is given by
\beq
v_n(p_T)= \frac{\delta J_1 + \delta J_2 +\delta J_3}{J_1^0+J_2^0} \frac{\epsilon_n}{2}\,,
\label{vnpt}
\eeq
and the integrated $v_n$ is
\beq
v_n= \frac{\int dp_T v_n(p_T) \frac{dN}{dY dp_T}}{\int dp_T \frac{dN}{dY dp_T}} = \frac{\int_0^\infty dp_T p_T (\delta J_1 + \delta J_2 +\delta J_3)}{\int_0^\infty dp_T p_T (J_1^0+J_2^0)} \frac{\epsilon_n}{2}\,.
\label{vn}
\eeq

\subsection{Small-$p_T$ region \label{secsmallpt}}

While it is straightforward to perform the remaining integral over $x_\perp$ (and also $p_T$) numerically, in this paper we are mainly interested in the analytic properties of $v_n$.
  They can be precisely studied in the small-$p_T$ region in which one can approximate the Bessel function as $I_n(z) \approx \frac{1}{n!} \left(\frac{z}{2}\right)^n$. Actually, since $z\sim p_T /TB^3$ and $B^3\gg 1$, the condition $z \lesssim 1$ is not very restrictive since it covers a wide region  $B^3T\gtrsim p_T$.
 In this region, we find
\beq
 J_1 &\approx & 4\pi m_T K_1(m_T/T)\int_0^\infty dx_\perp x_\perp \tau_0 \Biggl\{  1 -\beta_\pi
   +\frac{1}{n!}\frac{\delta \tau}{\tau_0}\left(\frac{p_T u_{\perp 0}}{2T}\right)^n\epsilon_n \cos n\phi_p \nn
&& \qquad \qquad + \frac{1-\beta_\pi}{n!}\left(\frac{p_T u_{\perp 0}}{2T}\right)^n \frac{n}{u_{\perp 0}}\left(\delta u_\perp  -\frac{\delta u_\phi}{x_\perp} \right)\epsilon_n\cos n\phi_p    \Biggr\}\,,
\eeq

\beq
J_2&\approx &-4\pi p_T K_0(m_T/T) \int_0^\infty dx_\perp x_\perp \tau_0
 \Biggl\{ \frac{\partial \tau_0}{\partial x_\perp} \frac{u_{\perp 0} p_T}{2T}\left(1-\widetilde{\beta}_\pi \right) + \frac{n}{n!2^n}\left(\frac{ u_{\perp 0} p_T}{T}\right)^{n-1}\frac{\partial \tau_0}{\partial x_\perp}\frac{\delta \tau}{\tau_0}
 \epsilon_n \cos n\phi_p  \nn  && \qquad + \frac{n}{n!2^n}\left(\frac{ u_{\perp 0} p_T}{T}\right)^{n-1}(1-\widetilde{\beta}_\pi)\left[\frac{\partial \tau_0}{\partial x_\perp}\frac{n-1}{u_{\perp 0}}\left(\delta u_\perp -\frac{\delta u_\phi}{x_\perp}\right)
   + \frac{\partial \delta \tau}{\partial x_\perp} \right]\epsilon_n \cos n\phi_p \Biggr\}\,,
\eeq
and
\beq
J_3\approx -4\pi p_T K_0(m_T/T)\int_0^\infty dx_\perp  \tau_0 \frac{n^2 \delta \tau}{2^n n!}\left(\frac{u_{\perp 0}p_T}{T}\right)^{n-1} (1-\widetilde{\beta}_\pi)\epsilon_n \cos n\phi_p\,.
\eeq

The remaining $x_\perp$-integrals can be done analytically. After straightforward, but  very tedious calculations, we arrive at

\beq
J_1^0 = 4\pi m_T K_1(m_T/T)\frac{16 L^3}{B^3} \left\{ 1- \frac{\kappa x_{\perp max}^2}{64L^2}\left(6+\frac{m_T^2}{2T^2}\frac{K_3-K_1}{K_1}-\frac{p_T^2}{T^2}\right) \right\}\,, \label{cut}
\eeq

\beq
J_2^0 = 4\pi K_0(m_T/T)\frac{2^{15}L^3p_T^2}{TB^9} \left\{\frac{1}{21} -\frac{\kappa}{640 }\left(12+\frac{m_T^2}{T^2}\frac{K_2 -K_0}{K_0}-\frac{p_T^2}{T^2}\right) \right\}\,,
\eeq

\beq
\delta J_1 &=&4\pi \frac{m_T}{T} K_1(m_T/T)\frac{\Gamma(3n)}{ \Gamma(4n)}
\frac{9\cdot 2^{6n}L^3 p_T^n}{B^{3(n+1)}T^{n-1}}  \nn &
\times& (n-1)\left\{
  \frac{2(3n+2)}{4n+1} -\frac{n\kappa}{8(3n-1)}\left( 6n+6+ \frac{m_T^2}{2T^2}\frac{K_3-K_1}{K_1}-\frac{p_T^2}{T^2}\right) \right\}\,, \label{deltaJ1}
  \eeq



\beq
\delta J_2 &=& 4\pi K_0(m_T/T)\frac{\Gamma(3n)}{\Gamma(4n)} \frac{9\cdot 2^{6n} L^3 p_T^n}{ B^{3(n+1)}T^{n-1}}
\nn
&&\times 2n\left\{\frac{6n^2-6n-5}{4n+1} -\frac{(6n^2-10n+1)\kappa}{48(3n-1)}\left( 6n+ \frac{m_T^2}{T^2}\frac{K_2 -K_0}{K_0}-\frac{p_T^2}{T^2}\right)\right\}\,, \label{deltaJ2}
\eeq

\beq
\delta J_3 &=& 4\pi  K_0(m_T/T) \frac{\Gamma(3n)}{\Gamma(4n)}\frac{9\cdot 2^{6n} L^3p_T^{n}}{  B^{3(n+1)}T^{n-1}} \nn
&& \qquad \times 2n\left\{1-\frac{(4n-1)\kappa}{48(3n-1)}\left(6n+
\frac{m_T^2}{T^2}\frac{K_2 -K_0}{K_0}-\frac{p_T^2}{T^2} \right)
\right\}\,,
\eeq
and
\beq
\delta J_2+\delta J_3 &=& 4\pi K_0(m_T/T)\frac{\Gamma(3n)}{\Gamma(4n)} \frac{9\cdot 2^{6n} L^3 p_T^n}{ B^{3(n+1)}T^{n-1}}
\nn
&\times& 2n(n-1)\left\{\frac{2(3n+2)}{4n+1} -\frac{n\kappa}{8(3n-1)}\left( 6n+\frac{m_T^2}{T^2}\frac{K_2 -K_0}{K_0}-\frac{p_T^2}{T^2}\right)\right\}\,. \label{deltaJ22}
\eeq
The $x_\perp$-integral in the viscous term of (\ref{cut}) is actually divergent and we cut it off at $x_{\perp max}$. This divergence is an artifact of Gubser's Navier-Stokes solution which becomes unphysical as $x_\perp \to \infty$ (the temperature becomes negative), and can be cured by including second-order hydrodynamic corrections \cite{Marrochio:2013wla}. The precise value of $x_{\perp max}$ will depend on such generalizations. Or phenomenologically, we prefer $x_{\perp max}\sim {\mathcal O}(L)$ (the nuclear diameter). In any case,  in the following, we simply ignore these cutoff-dependent terms in $J_1^0$ because they are independent of $n$, and are thus subleading compared with the viscous corrections from $\delta J_i$ which grow linearly in $n$ as we shall shortly see.

We now discuss the properties of $v_n(p_T)$ defined in (\ref{vnpt}). First, notice that
\beq
\frac{J_2^0}{J_1^0} \sim \frac{TK_0(m_T/T)}{m_T K_1(m_T/T)} \left(\frac{p_T}{B^3T}\right)^2 \ll 1\,, \label{large}
\eeq
for the unperturbed part.
As for the anisotropic part, $\delta J_1$ is dominant when $m_T \gg T$ and $\delta J_2$ is dominant when $n\gg 1$, while $\delta J_3$ is subleading.
  We thus find
  \beq
v_n(p_T) &\approx& \frac{\delta J_1}{J_1^0} \frac{\epsilon_n}{2} \nn
 &\approx& \frac{9\epsilon_n}{32} \frac{\Gamma(3n)}{\Gamma(4n)}
  \left(\frac{64 p_T}{B^3T}\right)^n (n-1)\Biggl\{
  \frac{2(3n+2)}{4n+1}   -\frac{n\kappa}{8(3n-1)}\left( 6n+6+ \frac{2m_T}{T}+3-\frac{p_T^2}{T^2}\right) \Biggr\}\,,
  \label{largem}
\eeq
    when $m_T \gg T$ and $n \sim {\mathcal O}(1)$, and
\beq
v_n(p_T) \approx \frac{\delta J_2}{J_1^0} \frac{\epsilon_n}{2} &\approx& \frac{27\epsilon_n}{32}  \frac{n(n-1)\Gamma(3n)}{\Gamma(4n)} \frac{TK_0(m_T/T)}{m_T K_1(m_T/T)}\left(\frac{64 p_T}{B^3T}\right)^n  \left(1-\frac{n\kappa}{6} \right) \nn
&\sim &  \epsilon_n n^2\left(\frac{27  p_T}{4B^3T}\right)^n e^{n-n\ln n} \left(1-\frac{n\kappa }{6}\right)\,,
\label{stir}
\eeq
 when $n\gg1$ and $m_T/T \sim {\mathcal O}(1)$.
In the second expression of (\ref{stir}), we used the Stirling's formula $\Gamma(n) \approx \sqrt{2\pi} e^{-n}n^{n-\frac{1}{2}}$ and neglected the $n$-independent factors.

We immediately notice the scaling relation $v_n(p_T)\propto p_T^n$. This follows trivially from the expansion of the Bessel function $I_n(z)$ and is a model-independent prediction of the Cooper-Frye formula. Another interesting feature is that the decrease of $v_n(p_T)$ at large-$n$ is faster than exponential $v_n(p_T) \sim 1/n^n$.
 Concerning the viscous corrections, we observe that the dependence on the Knudsen number is linear in $n$
  \beq
\label{sheark}
  \frac{v_n}{v_n^{ideal}} \sim 1-{\mathcal O}(n \kappa)\,.
  \eeq
  This is a generalization of the known behavior $v_2/v_2^{ideal} \sim 1-{\mathcal O}(\kappa)$ \cite{Bhalerao:2005mm,Hatta:2014upa}  to arbitrary $n$. This factor of $n$ arises because the power of $p_T$ is accompanied by a power of $u_{\perp 0}$ (see (\ref{u0})) so that $v_n \sim (p_T u_{\perp 0})^n \sim p_T^n (1-n\kappa)$. We also find that the contribution from $\delta f\sim \pi^{\mu\nu}$ in (\ref{non}) is suppressed by $1/n$ compared with that from the equilibrium distribution. However, we shall see later that the potentially large term $-p_T^2/T^2$ at high-$p_T$ brings about an interesting effect in the integrated $v_n$.

Note that in deriving the above formulas, we assumed $\epsilon_n\ll 1$ and kept only the linear terms in $\epsilon_n$. The results are thus more reliable in central collisions, while deviations are expected in peripheral collisions for which $\epsilon_n \sim {\mathcal O}(1)$.

\subsection{Large-$p_T$ region}

In the large-$p_T$ region the $x_\perp$-integral cannot be done exactly. However, in this region the Bessel function $I_n$ becomes independent of $n$
\beq
I_n(z) \approx \frac{e^z}{\sqrt{2\pi z}} \sim \exp\left(\frac{p_T}{T}\frac{2x_\perp (2L)^5}{B^3(x_\perp^2+L^2)^3}(1-\alpha) \right)\,.
\eeq
The $x_\perp$-integral can be evaluated by doing the saddle point at $x_\perp^*=L/\sqrt{5}$. The result is
\beq
\label{highpt}
v_n(p_T)
\approx \frac{\epsilon_n}{2}\frac{p_T}{T} \delta u_{\perp 0}^* = \epsilon_n \frac{500 p_T}{27TB^3}
\left(\frac{\sqrt{5}}{3}\right)^{n-1} \left(n-1-\frac{27\kappa}{200}n\right)   \,.
\eeq
We see that $v_n(p_T)$ rises linearly with $p_T$ and time $B^{-3}\sim \tau$, and gets suppressed by a uniform factor $v_n/v_{n-1}\sim \sqrt{5}/3 \approx 0.745$ as $n$ is increased. We also notice that, in contrast to the low-$p_T$ region, $v_n/v_n^{ideal}\sim 1-{\mathcal O}(\kappa)$ is independent of $n$.  Moreover, the contribution from $\delta f\sim \pi^{\mu\nu}$, which at first sight seems to be  important at high-$p_T$, actually cancels in the ratio $\delta J/J^0$.

\subsection{Integrated $v_n$}

Finally, we calculate the integrated $v_n$ in (\ref{vn}). We observe that, due to the $K$-Bessel function, the important region of the $p_T$-integral is $z \lesssim 1$, so that the results obtained in the small-$p_T$ region can be utilized.
Replacing the $K$-Bessel function with its asymptotic expression, we need to evaluate integrals of the form
\beq
 \int dp_T \exp\left(-\frac{\sqrt{m^2 +p_T^2}}{T} \right) p_T^{n+1}\,. \label{exa}
 \eeq
   For this purpose, we use the saddle point approximation assuming $n$ to be large, but not too large $B^3 \gtrsim n \gg 1$.
 First consider the heavy particle (`baryon') case $m\gg nT$.   There is a saddle point at $p_T^*\approx \sqrt{nmT}$.
  Evaluating around this saddle point, we find, from (\ref{deltaJ1}),
\beq
v_n \sim  \epsilon_n n^{3/2}\left(\frac{27}{4B^3}\sqrt{\frac{m}{T}}\right)^n \exp\left(\frac{n}{2}-\frac{n}{2}\ln n\right) \left(1-\frac{n\kappa}{36}\left(6-\frac{m}{T}\right)\right)\,.
\eeq
Compared with (\ref{stir}), the exponential factor has been halved.

Next consider the case $m\ll nT$ which in particular includes the massless limit $m=0$.
The saddle point is at $p_T^*=nT$ and this means that $\delta J_1$ and $\delta J_2$ are equally important to the integrated $v_n$
\beq
\frac{\delta J_2}{\delta J_1} \approx \frac{2nT}{p^*_T}\frac{K_0(p^*_T/T)}{K_1(p^*_T/T)} \approx 2\,. \label{2}
 \eeq
 Thus we can use either (\ref{deltaJ1}) or (\ref{deltaJ2}) to get
\beq
v_n \sim  \epsilon_n n^3\left(\frac{27}{4B^3}\right)^n \left(1-\frac{n\kappa}{36}\left(6-n\right)\right)\,. \label{light}
\eeq
In fact, when $m=0$, the $p_T$-integral with the full $K$-Bessel functions can be done exactly and we find, adding all the components $\delta J_{1,2,3}$,
\beq
v_n &\approx& \frac{9\epsilon_n}{32} \frac{\Gamma(3n)}{\Gamma(4n)}\left(\frac{128}{B^3}\right)^n \Gamma^2\left(\frac{n}{2}\right) \Biggl\{\frac{n^2(3n+2)^2(n-1)}{2(4n+1)} \nn && \qquad -\frac{n^3(n-1)\kappa}{16(3n-1)}\left(3(3n^2+3n+2) - \frac{n}{2}(n+2)(3n+4) \right) \Biggr\}\,, \label{from}
\eeq
 whose large-$n$ limit coincides with (\ref{light}).

We thus find that the $n$-dependence is rather different between the heavy and light particle cases.
   We also point out that, in both cases, the integrated $v_n$ inherits the property $v_n/v_n^{ideal} \sim 1-{\mathcal O}(n\kappa)$ of the unintegrated $v_n(p_T)$.  However, an interesting new feature is that the contribution from $\delta f\sim \pi^{\mu\nu}$ is large at the saddle point and this can flip  the sign of the $\kappa$-term when $n$ or $m/T$ is larger than some critical value.  In the present case with $\chi=1$, this occurs when $n> n^*\approx 3.8$ (from (\ref{from})), but the precise value of $n^*$ is a bit uncertain due to the cutoff-dependent term in (\ref{cut}) which enters the denominator of (\ref{vn}).\footnote{One might worry that the terms in (\ref{cut}) which derive from $\delta f\sim \pi^{\mu\nu}$ could become large after integrating over $p_T$. However, they are subleading because
\beq
\left.\int dp_T p_T J_1^0\, \right|_{\delta f} \sim \int dp_T p_T^2K_1(p_T/T)I_0(s) \beta_\pi \sim {\mathcal O}(B^{-6})\,.
\eeq
}
 For some choices of $\chi$ \cite{Dusling:2009df,Bhalerao:2013pza}, the $p_T$-dependence of $\delta f$ is weaker and  $n^*$ may not exist.

 (\ref{light}) exhibits the scaling $v_n\sim B^{-3n} \sim \tau^n$ with respect to the freezeout time. In particular, when $n=2$, the above result for $v_2$ parametrically agrees with the quantity
\beq
\epsilon_p(\tau) \equiv
\frac{\int d^2x_\perp (T_{11}-T_{22}) }{\int d^2x_\perp ( T_{11}+T_{22})}\sim \epsilon_2\frac{\tau^2}{L^2}\left(1-{\mathcal O}(\kappa)\right)\,, \label{ep}
\eeq
computed in Ref.~\cite{Hatta:2014upa}. It has been found numerically  \cite{Kolb:1999it} that $\epsilon_p$ is a good measure of $v_2$ in that they are monotonously related to each other as the freezeout time is varied.
   We have thus analytically confirmed the proposed connection between $\epsilon_p$ and $v_2$ at least in the early time regime. They indeed have the same parametric form with respect to $\tau$ and the shear viscosity.

 As a side remark, the linear $n$-dependence in the viscous correction derived here is in contrast to the formula proposed in  \cite{Staig:2011wj,Lacey:2013is} which reads, in the present notation
\beq
\frac{v_n}{\epsilon_n} \sim  \exp\left(-\frac{2n^2}{3LT}\frac{\eta}{S}\right) \sim   1-{\mathcal O}(n^2\kappa)\,.
\label{edshuryak}
\eeq
Ref.\ \cite{Lacey:2013is} has used the formula above to fit the $n$-dependence of LHC flow data and performed a phenomenological estimate of the viscous damping induced by $\eta/S$ on the flow coefficients. However, we note that such a formula is in disagreement with the analytical study performed in this paper and, moreover, (\ref{edshuryak}) may not be valid in event-by-event simulations, as recently pointed out in Ref.\ \cite{Gorda:2014msa}.

\section{Mixing between $v_n$ and $v_{2n}$}

In this section, we briefly discuss the mixing of different harmonics based on this approximate analytical solution. See Ref.~\cite{Teaney:2012ke, Floerchinger:2013rya} and follow up works for more comprehensive studies in this regard. Here we only consider the mixing of $v_n$ and $v_{2n}$, but other types of mixing, such as that of even and odd harmonics, can be studied similarly.  For simplicity, consider the inviscid case $\kappa=0$ and keep only $J_1$. Expanding the Boltzmann factor to linear order in $\epsilon_{2n}$ and quadratic order in $\epsilon_n$, we get
\beq
J_1&=& 2m_T K_1(m_T/T)\int dx_\perp x_\perp \tau_0 \int d\phi e^{\frac{p_T u_{\perp 0}}{T}\cos (\phi-\phi_p)} \nn && \times \Biggl\{ 1+ \frac{p_T}{T}\left(\delta u_{\perp}^{(2n)}\cos 2n\phi \cos (\phi-\phi_p)-\frac{1}{x_\perp}\delta u_\phi^{(2n)} \sin 2n\phi \sin (\phi-\phi_p) \right)\epsilon_{2n} \nn
&& \qquad  + \frac{p_T^2}{2T^2}\left(\delta u_{\perp}^{(n)}\cos n\phi \cos (\phi-\phi_p)-\frac{1 }{x_\perp}\delta u_\phi^{(n)} \sin n\phi \sin (\phi-\phi_p) \right)^2\epsilon_n^2 + \cdots \Biggr\} \nn
&=& 4\pi m_T K_1(m_T/T)\int dx_\perp x_\perp \tau_0 \Biggl\{I_0(z) +\frac{p_T}{T}\left( \delta u_{\perp}^{(2n)}I'_{2n}(z) -  \frac{\delta u_\phi^{(2n)}}{x_\perp}\frac{2n I_{2n}(z)}{z} \right) \epsilon_{2n}\cos 2n\phi_p \nn &&
+\frac{p_T^2}{4T^2} \left( (\delta u_\perp^{(n)})^2 I''_{2n}(z) -\frac{\delta u_\perp^{(n)} \delta u_\phi^{(n)}}{x_\perp}\left(\frac{4n I_{2n}(z)}{z}\right)' -\frac{(\delta u_\phi^{(n)})^2}{x_\perp^2}(I_{2n}(z) - I''_{2n}(z))\right)\epsilon_n^2 \cos 2n\phi_p
\Biggr\}\,.
\eeq

We again consider the small-$p_T$ and large-$p_T$ regions separately. \\

(i) Small-$p_T$ region: Expanding the $I$-Bessel functions as before, we find
\beq
J_1 &\approx& 4\pi m_T K_1(m_T/T)\int dx_\perp x_\perp \tau_0 \Biggl\{1 +\left(\frac{p_T u_{\perp 0}}{2T}\right)^{2n} \Biggl[ \frac{1}{(2n-1)! u_{\perp 0}}\left( \delta u_{\perp}^{(2n)} -  \frac{\delta u_\phi^{(2n)}}{x_\perp} \right) \epsilon_{2n} \nn
&& \qquad + \frac{1}{4(2n-2)! u_{\perp 0}^2}\left( \delta u_\perp^{(n)} -\frac{ \delta u_\phi^{(n)}}{x_\perp}\right)^2  \epsilon_n^2 \Biggr]\cos 2n\phi_p
\Biggr\} \nn
&\approx& 4\pi m_T K_1(m_T/T) \frac{16L^3}{B^3} \left\{1+\frac{27n}{16}\left(\frac{64 p_T}{B^3T}\right)^{2n}\frac{\Gamma(6n)}{\Gamma(8n)} \left(\epsilon_{2n} + \frac{9}{8}n^2 \epsilon_n^2\right) \cos 2n \phi_p \right\}\,,
\eeq
 where we kept only the leading term in $n$. We see that the contamination from the lower order harmonics has the same
  power-law in $p_T$, and effectively shifts the eccentricity as
 \beq
 \epsilon_{2n} \to \epsilon_{2n} + {\mathcal O}(n^2\epsilon_n^2)\,. \label{shift}
 \eeq
(ii) Large-$p_T$ region: The $I$-Bessel functions become independent of $n$.
\beq
J_1
 \approx 4\pi m_T K_1(m_T/T)\int dx_\perp x_\perp \tau_0 I_0(z)\Biggl\{1 +\frac{p_T}{T} \delta u_{\perp}^{(2n)}  \epsilon_{2n}\cos 2n\phi_p
+\frac{p_T^2}{4T^2}(\delta u_\perp^{(n)})^2 \epsilon_n^2 \cos 2n\phi_p + \cdots
\Biggr\}\,.  \nonumber
\eeq
Doing the saddle point, we find
\beq
v_{2n}(p_T) \approx \frac{p_T}{2TB^3}\left(\frac{10}{3}\right)^3\left(\frac{\sqrt{5}}{3}\right)^{2n-1} (2n-1) \epsilon_{2n} + \frac{1}{2}\left(\frac{p_T}{2TB^3}\right)^2\left(\frac{10}{3}\right)^6
\left(\frac{\sqrt{5}}{3}\right)^{2n-2} (n-1)^2 \epsilon_n^2\,. \label{v2npt}
\eeq
We notice that the second term is just $\frac{1}{2}v_n^2(p_T)$ \cite{Borghini:2005kd} and becomes dominant at sufficiently high-$p_T$. \\

(iii) Integrated $v_{2n}$:
Since the $p_T$-integral is dominated by the low-$p_T$ region, the integrated $v_{2n}$ is modified by the same shift (\ref{shift}), namely,
 \beq
 v_{2n} \to v_{2n} \frac{\epsilon_{2n} + {\mathcal O}(n^2\epsilon_n^2)}{\epsilon_{2n}}\,.
 \label{v2ntot}
 \eeq
 For $n=2$, the correction is negligible for central collisions where $\epsilon_2 \sim \epsilon_4 \ll 1$, but it will be important for peripheral collisions. Note also that the mixing is enhanced for large values of $n$ by a factor $n^2$.

\section{Bulk viscosity}

In this section, we study the effect of bulk viscosity on $v_n$ which has attracted some attention lately \cite{Monnai:2009ad,Dusling:2011fd,Noronha-Hostler:2013gga,Noronha-Hostler:2014dqa}. Of course, since the bulk viscosity is absent in the presence of strict conformal symmetry, such a study is necessarily of approximate nature, largely motivated by phenomenological interest. Nevertheless, by treating the bulk viscous effect as small perturbation, we shall draw some useful observations. In this section we neglect shear viscosity effects, i.e., $\kappa=0$.

As in the shear case, the bulk viscosity affects $v_n$ in two ways. Firstly, it can modify the freezeout surface and hence the flow velocity on that surface. Secondly, it induces a new term $\delta f^{bulk}$ in the distribution function. Here we work in the `probe approximation' and consider only the latter effect.
 The backreaction of bulk effects on the flow profile is harder to implement in the present setup, and we leave it to future work.

For a single component gas in the small mass limit $m\ll T$, within the 14-moment approximation the bulk viscous correction to the particle distribution may be written as an expansion in powers of $(u_\mu p^\mu)$ given by \cite{Denicol:2014vaa}
\beq
\frac{\delta f^{bulk}}{f_{eq}}=\frac{12T^2}{m^2}\left[12+\frac{8}{T}u_\mu p^\mu + \frac{1}{T^2}(u_\mu p^\mu)^2\right]\frac{ \nabla_\mu u^\mu}{T}\frac{\zeta}{S}\,,
\label{deni}
\eeq
 where we take the bulk viscosity-to-entropy ratio $\zeta/S$ as a free (small) parameter and have assumed the Navier-Stokes relation between the bulk scalar $\Pi$ and $\nabla_\mu u^\mu$.
 Let us first calculate the correction to $J_1$ induced by (\ref{deni}). In the early time regime $\tau \ll L$, we may approximate  $\nabla_\mu u^\mu \approx 1/\tau$ and get
\beq
J_1^{bulk}&=& \overline{\zeta}\frac{m_T}{T}\int dx_\perp dyd\phi\, x_\perp \cosh(y-Y)e^{U+\epsilon_n\delta U}  \left(12+8(U+\epsilon_n\delta U) + (U+\epsilon_n\delta U)^2\right)\nn
&\approx& \overline{\zeta} \frac{m_T}{T}\int dx_\perp dyd\phi\, x_\perp  \cosh(y-Y)e^{U} \left(12+8U+U^2 +\left(20+10U+ U^2\right)\epsilon_n \delta U \right)\,,
\eeq
where we introduced the  notation $\overline{\zeta}\equiv \frac{12T^2}{m^2}\frac{\zeta}{S}$ and abbreviated
\beq
\frac{u^\mu p_\mu}{T} &=& \frac{1}{T}\Biggl[-m_T\cosh(y-Y) + p_Tu_{\perp 0} \cos (\phi-\phi_p)  \nn
&& +p_T\epsilon_n\left(\delta u_\perp \cos (\phi-\phi_p)
\cos n\phi -\frac{\delta u_\phi}{x_\perp}\sin (\phi-\phi_p)\sin n\phi\right) \Biggr] \equiv U+\epsilon_n\delta U\,.
\eeq
In order to perform the $y$, $\phi$-integrals efficiently, we introduce the following trick.  We first evaluate
\beq
X_1(a) &=& \int dyd\phi\,   \cosh(y-Y)e^{a U} =4\pi K_1(am_T/T)I_0(az) \approx 4\pi K_1(am_T/T) \,, \quad (z\ll 1)  \nn
X_2(a)&=&\int  dyd\phi\,   \cosh(y-Y)e^{a U}  \delta U \nn
 &=& 4\pi K_1(am_T/T)\frac{p_T}{2T}\left(\left(\delta u_\perp -\frac{\delta u_\phi}{x_\perp}\right)I_{n-1}(az)+\left(\delta u_\perp + \frac{\delta u_\phi}{x_\perp}\right)I_{n+1}(az)\right)\cos n\phi_p \nn
  &\approx & 4\pi K_1(am_T/T)\frac{p_T}{T}\frac{(az)^{n-1}}{2^n(n-1)!}\left(\delta u_\perp -\frac{\delta u_\phi}{x_\perp}\right)\cos n\phi_p \,, \quad (z\ll 1)
\eeq
 where $a$ is an auxiliary parameter. We can then write, for $z\lesssim 1$,
\beq
J_1^{bulk}&=&\overline{\zeta} \frac{m_T}{T}\left.\int dx_\perp x_\perp \left(12X_1 +8X'_1+X''_1 +\epsilon_n \left(20 X_2+10X'_2+ X''_2\right) \right)\right|_{a=1} \nn
 &=& 4\pi \overline{\zeta} \frac{m_T}{T}\int dx_\perp x_\perp  \Biggl[12K_1 -\frac{4m_T}{T}(K_0+K_2)+\frac{m_T^2}{4T^2}
 (3K_1+K_3)\nn &&
 \qquad \qquad +\frac{p_T}{T}\left( (n+3)(n+4)K_1-\frac{m_T}{T}(n+4)(K_0+K_2) + \frac{m_T^2}{4T^2}(3K_1+K_3)\right) \nn && \qquad \qquad \qquad \times\frac{z^{n-1}}{2^n (n-1)!}\left(\delta u_\perp -\frac{\delta u_\phi}{x_\perp}\right) \epsilon_n \cos n\phi_p \Biggr] \nn
 &=& J_1^{bulk,0} + \delta J_1^{bulk} \epsilon_n \cos n\phi_p\,.
 \eeq
 The remaining $x_\perp$-integral can be performed straightforwardly and we get
 \beq
J_1^{bulk,0} &=&  2\pi \overline{\zeta}  x_{max}^2 \frac{m_T}{T}  \left(12K_1 -\frac{4m_T}{T}(K_0+K_2)+\frac{m_T^2}{4T^2}
 (3K_1+K_3) \right)\,, \nn
   \delta J_1^{bulk}&=&9\pi \overline{\zeta} L^2 \frac{m_T}{T}\left( (n+3)(n+4)K_1-\frac{m_T}{T}(n+4)(K_0+K_2) + \frac{m_T^2}{4T^2}(3K_1+K_3)\right) \nn
    &&  \qquad \qquad \times\left(\frac{64p_T}{B^3T}\right)^n
  \frac{n(n-1)\Gamma(3n)}{(3n-1)\Gamma(4n)}\,.
\eeq
In  these results, all the terms in (\ref{deni}) are equally important. In particular, when $n$ or $m_T/T$ is large, the last term $(u_\mu p^\mu)^2$ gives the dominant contribution.
Similarly, we find
\beq
\delta J_2^{bulk}+\delta J_3^{bulk}&=&18\pi \overline{\zeta} L^2 \left( (n+2)(n+3)K_0-2(n+3)\frac{m_T}{T}K_1 + \frac{m_T^2}{2T^2}(K_0+K_2)\right) \nn
    &&  \qquad \qquad \times\left(\frac{64p_T}{B^3T}\right)^n
  \frac{n^2(n-1)\Gamma(3n)}{(3n-1)\Gamma(4n)}\,.
\eeq

As before, we can use this small-$p_T$ result to estimate the correction to the integrated $v_n$.
However, there is a caveat. If one naively approximates $m_T\approx p_T$, then the integrals $\int dp_T p_T  J_1^{bulk,0}$ and $\int dp_T p_T \delta J_i^{bulk}$ vanish exactly. Therefore, one has  to first expand the integrand to ${\mathcal O}(m^2)$ and then integrate.\footnote{We have actually neglected the ${\mathcal O}(B^{-6})$ terms which come from the expansion of the $I$-Bessel functions. This means that, strictly speaking, the present analysis is valid only for $B^3 \gg T/m\gg 1$. }  The result is
\beq
\int dp_T p_T J_1^{bulk,0} = -6\pi \overline{\zeta} x_{max}^2 m^2\,,
\eeq
\beq
\int dp_T p_T\delta J_1^{bulk} = -\frac{27\pi}{4} \overline{\zeta} m^2 L^2 \left(\frac{128}{B^3}\right)^n
  \frac{n^3(n-1)\Gamma(3n)}{(3n-1)\Gamma(4n)} \Gamma^2\left(\frac{n}{2}\right)\,,
\eeq
\beq
\int dp_T p_T(\delta J_2^{bulk}+\delta J_3^{bulk}) &=& -\frac{54\pi}{4} \overline{\zeta} m^2L^2 \left(\frac{128}{B^3}\right)^n
  \frac{n^3(n-1)\Gamma(3n)}{(3n-1)\Gamma(4n)}  \Gamma^2\left(\frac{n}{2}\right)\,.
\eeq
We again find that $\delta J_2\approx 2\delta J_1$, see (\ref{2}). Adding these terms, we obtain
 \beq
 \delta v_n^{bulk} &\approx& \frac{81}{128}\left(\frac{128}{B^3}\right)^{n}
 \frac{n^2(n-1)\Gamma(3n)}{\Gamma(4n)} \Gamma^2\left(\frac{n}{2}\right)
 \left(\frac{(3n+2)^2}{4(4n+1)}\frac{x_{max}^2}{L^2} -\frac{3n}{3n-1}\right)
 \frac{B^2\zeta}{CS}\epsilon_n\,,
  \label{last}
 \eeq
  where the first and second terms in the brackets come from the denominator and numerator of (\ref{vn}), respectively.
 Assuming $x_{max}\sim {\mathcal O}(L)$ (see a comment after (\ref{deltaJ22})), we find
  \begin{equation}
\label{bulkk}
\frac{v_n}{v_n^{ideal}}-1
\sim {\mathcal O}(\kappa^{bulk})-{\mathcal O}(\kappa^{bulk}/n)\,,
\end{equation}
 where the Knudsen number associated with the bulk viscosity is (cf. (\ref{kn}))
  \beq
  \kappa^{bulk} \sim \frac{B^2}{C}\frac{\zeta}{S}\,.
  \eeq
We see that, in contrast to the shear case (\ref{light}), we no longer have an enhancement by a positive power of $n$. Rather, $\delta J^{bulk}$ contribution is suppressed by a factor $1/n$, and because of this, the cutoff-dependent term $J_{1}^{bulk,0}$ cannot be neglected. The sign of the right hand side of (\ref{bulkk}) is uncertain and model-dependent, though we can say it is positive for sufficiently  large values of $n$.

 There is actually a bigger source of uncertainty in comparing our results with numerical simulations. Although the simple Knudsen number scaling (\ref{bulkk})
  parallels what we have found in the shear case, the result crucially depends on  a  delicate cancelation among the three terms in (\ref{deni}). This may not occur in  some approximations in which the relative weight of these terms is different.\footnote{For instance, in the explicitly non-conformal numerical calculations of \cite{Noronha-Hostler:2013gga,Noronha-Hostler:2014dqa}, the small $m/T$ limit was not assumed to obtain the coefficients in the bulk correction to the distribution function at freezeout. These coefficients are in fact very different than those used here.}
  It could also be spoiled by numerical errors after performing complicated integrals over $x_\perp$ and $\phi$. In such circumstances, the integral of $\delta J^{bulk}$ would be parametrically larger than (\ref{bulkk}) by a factor $n^3 T^2/m^2\gg 1$
\beq
\frac{v_n}{v_n^{ideal}}-1 \sim \pm n^2\frac{T^2}{m^2}\kappa^{bulk}\,,
\eeq
 and the sign can be either positive or negative.

Our analysis thus suggests that some care is needed when interpreting the result for $\delta v_n^{bulk}$ found in recent  numerical simulations.  The final value of $\delta v_n^{bulk}$, even its sign, can strongly depend on the details of the models, the freezeout time as well as the precision of numerics.

\section{Scaling Phenomenology \label{secpheno}}
To compare our results for $v_n$ to experimental data, we need to fix the constants $C$ and $B$ in terms of  bulk
observables. For this purpose, let us compute $dN/dY$ and the average transverse momentum $\langle p_T\rangle$ within the same framework. Since these observables are dominated by the ideal, isotropic part of the flow, the calculation is very simple and we obtain
 \beq
\frac{dN}{dY} &=& \frac{1}{(2\pi)^2}\int dp_T p_T (J_1^0+J_2^0) \approx \frac{4C^3}{\pi}
\,,  \label{dndy}
\eeq

\begin{equation}
\ave{p_T} \equiv \left( \frac{dN}{dY} \right)^{-1}  \int p_T dp_T \frac{dN}{ dYdp_T  }\approx \frac{3\pi T}{4}=\frac{3\pi CB}{8L}
\,. \label{pt}
\end{equation}
   From (\ref{dndy}) and (\ref{pt}), we find
\beq
C \sim \left(\frac{dN}{dY}\right)^{1/3}\,, \qquad \frac{1}{B^3} \sim \frac{1}{\langle p_T\rangle^3L^3}\frac{dN}{dY}\,. \label{bb}
\eeq
  (\ref{bb}) leads to the following scaling relations in terms of observables:
\beq
\left(\frac{v_n(p_T)}{\epsilon_n}\right)^{1/n}  \sim \frac{p_T}{A_\perp^{3/2}\langle p_T\rangle^4}\frac{dN}{dY}(1- n \kappa)\,,  \qquad \left(\frac{v_n}{\epsilon_n}\right)^{1/n} \sim  \frac{1}{A_\perp^{3/2}\langle p_T\rangle^3}\frac{dN}{dY}(1-n \kappa)\,, \label{scalingnpt}
\eeq
\beq
\kappa \sim \frac{B^2}{C}\frac{\eta}{S} \sim \frac{A_\perp\langle p_T\rangle^2}{dN/dY} \frac{\eta}{S}\,,
\label{kappascale}
\eeq
 where $A_\perp \sim L^2$ is the overlapping area (note that  $A_\perp^{3/2}\propto N_{part}$, the number of participants).
 One may also include in (\ref{scalingnpt}) the contributions from $\delta f^{shear}$ and $\delta f^{bulk}$. However, as we discussed already, the $n$-dependence of these terms can be strongly model-dependent.


 Finally, the dependence on the collision energy $\sqrt{s}$ and $N_{part}$ can be deduced by using the following empirical  formulae \cite{whitephobos,alicemult,alicept,gtscaling}
\begin{equation}
\frac{dN}{dY} \sim N_{part} (\sqrt{s})^\gamma\,,  \phantom{AA} \qquad  \ave{p_T} \sim F \left( \frac{1}{N_{part}^{2/3}}\frac{dN}{dY}  \right) \sim F\left(N_{part}^{1/3} (\sqrt{s})^\gamma\right)\,, \phantom{AA} \label{aa}
\end{equation}
 where $\gamma\approx 0.15$ in $AA$ collisions and $\gamma\approx 0.1$  in $pA$ and $pp$ collisions, and $F$ is a  rising function of its argument (see, Ref.~\cite{alicept}). Using this we arrive at

\beq
\left(\frac{v_n}{\epsilon_n}\right)^{1/n} \sim  (\sqrt{s})^\gamma G\left(N_{part}^{1/3} (\sqrt{s})^\gamma\right) (1-n \kappa)\,, \qquad \kappa \sim H\left(N_{part}^{1/3} (\sqrt{s})^\gamma\right)\frac{\eta}{S}\,, \label{scalingexp}
\eeq
 where $G(x)=F^{-3}(x)$ and $H(x) = F^2(x)/x$.

It is worth spending a few words to what extent do these scalings agree with heavy-ion systematics as a whole, measured for all energies and system sizes.
We note that, up to $\sim {\mathcal O}(\kappa)$, the overall $v_n/\epsilon_n$ dependence of Eq.~\eqref{scalingnpt} differs from what is  expected from the scale invariance of nearly-ideal hydrodynamics  \cite{Bhalerao:2005mm}
\begin{equation}
\label{v2const}
\frac{v_2}{\epsilon_2} \sim const. - \order{\kappa}
\end{equation}
 and also from the more popular multiplicity scaling
\beq
\frac{v_2}{\epsilon_2} \sim f\left(\frac{1}{A_\perp}\frac{dN}{dY}\right)\,,
\label{multi}
\eeq
 which is observed  in experimental data (see Fig.~24 of \cite{cms}, Fig.~10 of \cite{scale} and Fig.~4 of \cite{paridge1}).

It is actually easy to reconcile our result with Eq. \eqref{v2const} by naively extrapolating $B^3 \to 1$ or $\tau \to  L$, arguing that $\tau\sim {\mathcal O}(L)$ is the typical build-up time of $v_n$'s. However, working in the regime $B^3>1$  allows us to uncover a `hidden' scaling variable (\ref{scalingnpt}), and this scaling, being  solely in terms of observables, could be preserved up to the realistic freezeout time.    Moreover, as Eq. \eqref{kappascale} shows, $\kappa \sim A_\perp (dN/dy)^{-1}$;  Hence, the scaling in Eq. \eqref{multi} and the scaling in Eq. \eqref{v2const} are compatible if the leading dependence of multiplicity and size on $v_2/\epsilon$ comes from $\kappa$.


Numerical simulations \cite{hsong} seem to confirm this, but Eq. \eqref{scalingnpt} givs a different dependence from Eq. \eqref{multi}.
This discrepancy is more worrisome. We note, however, that artificially introducing an extra factor $A_\perp^{1/2}\sim L$ in the numerator of \eqref{scalingnpt} or (\ref{scalingexp}) would precisely match the scaling (\ref{multi}). It is not clear whether such a factor is dynamically generated as we go beyond the early-time approximation.\footnote{The time scale for the shock-rarefaction wave to cross the system is precisely of that order, although given the non-monotonicity of the development of flow harmonics with respect to  time, it is not guaranteed to occur. }
Given the availabilitiy of experimental data scanning across energies, system sizes and $p_T$, the scaling of \eqref{scalingnpt} or (\ref{scalingexp}) could be compared with experimental data.    If it works, it might lead to a way of obtaining the parameters characterizing the system, such as $\kappa$, the initial temperature and the lifetime, independently of hydrodynamic simulations and fits to data.


 The dependence of measured $v_2(p_T)$  on initial temperature between classes of events with the same geometry but different $\sqrt{s}$ appears to be  weaker than Eqs.~\eqref{largem} and \eqref{scalingexp} would suggest (see Fig.~9 of \cite{scanv2paper}).  Note that more sophisticated hydrodynamic calculations have the same problem \cite{kestin}, since $v_2(p_T)$, unlike the data, decreases in each $p_T$ bin with the initial temperature.   It will therefore be interesting to see how far experimental scaling plots will remain viable when more $v_n$, energies and system sizes are compared.

Concerning the $p_T$-dependence, LHC results such as \cite{alicevn,aaridgeatlas} can be used to check the scaling of $v_n(p_T)$ with respect to $n$ and $p_T$ (Eq.~(\ref{scalingnpt})) within similar initial temperature and system sizes. 
Since this scaling is a direct result of the $I_n(z) \simeq z^n/2^n n!$ approximation in section \ref{secsmallpt}, and $I_n$ will show up in any integration over an approximately azimuthally symmetric fireball, we expect it could persist within more realistic calculations \cite{heinzvn,hsong}.
Furthermore, the difference between Eq.~(\ref{scalingnpt}) and the intermediate $p_T$ limit described in Eq.~(\ref{highpt}), together with higher-$p_T$ $v_{2,3}$ data \cite{cms,highptscaling}, can then be used to estimate at what $p_T$ does the Knudsen number $\kappa$ become non-negligible and the hydrodynamic approximation breaks down.

Comparing $v_{2,3}$ results within systems of similar multiplicity but different sizes ($pA,dA,AA$ collisions at RHIC and LHC \cite{paridge1,paridge2,scale}) can be used to test whether the system response to changes in $A_\perp$ follows Eq.~(\ref{scalingnpt}).
Scans in $\sqrt{s}$ and system size at RHIC \cite{scanv2paper} can be used to comprehensively test the equations of the previous section against changes in {\em both} initial parton density and size.
Once these scalings are established for the leading term in $v_2$,  a precise measurement of how $v_n$ decays with $n$ can be used to disentangle bulk and shear viscosity contributions (though this requires a precise knowledge of $\delta f^{shear}$ and $\delta f^{bulk}$).

The $v_n$'s calculated here can also be summed into a two-particle azimuthal correlation which, given the  is given by \cite{keanestock}
\begin{equation}
\frac{d N}{dp_{T1}dp_{T2} d(\phi_1 -\phi_2)} \sim \sum_{n} v_n\left( p_{T1} \right) v_n\left( p_{T2} \right)  \cos\left( n \left( \phi_1-\phi_2 \right) \right) \,.
\end{equation}
This correlation function would by its nature include, on the same setting, the Knudsen number corrections as well as the freezeout-driven $p_T$-dependent mixing found in Eqs \eqref{v2ntot} and \eqref{v2npt}.  A similar mixing was found in more realistic numerical simulations \cite{factor} but needs sufficient precision control over initial conditions, including the dependence of eccentricity on color coherence \cite{raju} and multi-nucleon correlations \cite{bron}, to be studied quantitatively.

\section{Discussion and Conclusions}

In this paper we have analytically computed the anisotropic flow harmonic coefficients $v_n$ associated with perturbations around Gubser flow in the Navier-Stokes approximation of fluid dynamics. This was possible due to the observation that the hydrodynamic perturbation equations derived in \cite{Gubser:2010ui} can be treated analytically at early times.

 Within this model, we have been able to explicitly derive the $n$-dependence of  $v_n$, for both unintegrated and integrated in $p_T$. In particular, we find that the suppression at large-$n$ is faster than exponential $v_n(p_T) \sim e^{-n\ln n}$ in the unintegrated case (\ref{stir}), and exponential $v_n\sim e^{-n \ln B^3}$ in the integrated case (\ref{light}), respectively. Moreover, they obey the relation $v_n/v_n^{ideal} \sim 1-\mathcal{O}(n \,\eta/S)$, which nicely encodes the dependence on $\eta/S$, and is a direct generalization of the previously found behavior for elliptic ($n=2$) flow \cite{Bhalerao:2005mm,Hatta:2014upa} to arbitrary $n$. Our calculations also allowed us to confirm, in an analytical manner, the connection between the integrated elliptic flow (including shear viscous corrections) and the spatial anisotropy $\epsilon_p$ previously studied numerically in \cite{Kolb:1999it} and analytically in \cite{Hatta:2014upa}. Furthermore, it was possible to compute the mixing between $v_n$ and $v_{2n}$ and find simple relations between them. For the integrated coefficients,
 this mixing becomes important only for peripheral collisions.

While we found the generally expected result that shear viscosity decreases the value of flow coefficients (apart from a subtlety regarding the contribution from the $\delta f \sim \pi^{\mu\nu}$ term), the result for the bulk viscosity remains inconclusive and cannot be directly compared with the numerical calculations performed in \cite{Noronha-Hostler:2013gga,Noronha-Hostler:2014dqa}.
 Rather, our results suggest that even the overall sign of the bulk viscosity correction may be extremely sensitive to model assumptions. This points towards a more detailed calculation of the bulk viscous corrections to the particle distribution currently used in hydrodynamic simulations.

Regarding the interpretation of our results with respect to the phenomenology of heavy-ion collisions, even though the formulae presented here can be used in quantitative fits, we urge caution in any physical interpretation of the parameters, as the present model is undoubtedly highly simplified. The initial distribution is not even qualitatively similar to a Glauber superposition of sharp nucleon-centered energy density peaks. The equation of state and transport coefficients are constrained by conformal invariance, and therefore cannot incorporate the characteristic $T_c$ scale of QCD. The interplay of resonance decay and jets with collective flow is not taken into account \cite{Andrade:2014swa}.
And finally, hydrodynamic simulations tuned to particle spectra point to a typical lifetime which is 1$\sim$1.5 times the size of the system \cite{kolb}, while the approximations made in the current work do not apply unless freezeout is parametrically shorter than the system size.

For a qualitative analysis of the type described in the conclusion of section \ref{secpheno}, however,  these may not be major obstacles.  The shorter lifetime of the system\footnote{which, incidentally, is what is indicated by HBT data \cite{kolb} so this approximation might not be so bad, although it would be surprising if freezeout time was always parametrically smaller than $L$}  and nearly conformal EoS and transport may not affect momentum-anisotropies, which are expected to be formed predominantly in the early high temperature evolution of the system (bulk viscosity may be an exception to this, however). And, if dimensional analysis is good enough for parametric estimates, a unrealistic geometry may not be so important in an investigation of the {\em response} to changes in system size, $p_T$, density and flow harmonic number.

Hence, Section \ref{secpheno} shows that our analytical expressions provide us quite straightforward links to experimental observables ($dN/dy$, $\ave{p_T}$, and $v_n/\epsilon_n$). 
Given that the data collected in the last few years can be scanned in many different ways (across $\sqrt{s}$, $p_T$, $N_{part}$, $dN/dy$, harmonic number, system size and so on), such a simple dependence can be thoroughly tested on a qualitative level.
On the theoretical side, the scalings described here can be tested by scanning the codes used in \cite{pratt,heinz,bass} in system size $L$ and the freezeout temperature $T\sim CB/L$, and by checking to what extent can equations such as \eqref{largem} be reproduced in a realistic hydrodynamic simulation.
If it turns out they can, it would confirm that the results derived in the previous section can be used for phenomenology despite the simplifications made in our model.    If they cannot, numerical simulations might yield similar, but more phenomenologically applicable, scaling relations.

This analysis does not require a quantitative description of the data, but merely a smooth variation of experimental observables across all scanning variables, something which seems to occur from lower energy RHIC to LHC energies. However, Eq.~(\ref{scalingnpt}) makes it clear that the response of azimuthal observables to changes of $n,p_T,L$ as well as $dN/dy,\ave{p_T}$ is highly constraining, and can therefore be experimentally tested. 
If the model prediction survives such a qualitative test, an extraction of the physically meaningful $B$ and $\kappa$ parameters is straightforward, and changes in $\kappa$ and $P/T^4,\eta,\zeta$, expected from fundamental QCD, can be studied.
In particular, any scaling breakdowns due to the change in $P/T^4$ or $\eta/s$ would turn up in the scaling formulae summarized in the previous section.


In conclusion, in this paper we provided analytical expressions for the anisotropic flow coefficients which contain all the characteristics, in simplified form, of a ``realistic'' heavy-ion calculation: longitudinal and azimuthally anisotropic flow, viscous evolution, and particle production via isothermal freezeout. We have linked the parameters of this model to experimental observables, both bulk event characteristics (the multiplicity and average transverse momentum) and the response of the system to geometric anisotropies at different Fourier harmonics and $p_T$ bins. While this calculation is simplistic in comparison to current numerical simulations, we hope it can lead to a more straightforward understanding of how flow anisotropies depend on the underlying event structure, both in experimental data and numerical codes.

\section*{Acknowledgements}
We thank  G.~S.~Denicol and U.~Heinz for useful remarks. J.~N.\ thanks the Conselho Nacional de Desenvolvimento Cient\'ifico e Tecnol\'ogico (CNPq) and Funda\c c\~ao de Amparo \`a Pesquisa do Estado de S\~ao Paulo (FAPESP) for support. G.~T.\ thanks the Funda\c c\~ao de Amparo \`a Pesquisa do Estado de S\~ao Paulo (FAPESP) for
support. B.~X.
wishes to thank Dr. F. Yuan and the nuclear theory group at Lawrence Berkeley National Laboratory for
hospitality and support during his visit when this work is finalized.


\begin{thebibliography}{99}


\bibitem{sqgpmiklos}
  M.~Gyulassy and L.~McLerran,
  Nucl.\ Phys.\  A {\bf 750}, 30 (2005).

\bibitem{sqgpshuryak}
  E.~Shuryak,
  Prog.\ Part.\ Nucl.\ Phys.\  {\bf 53}, 273 (2004).

\bibitem{hwa}
  R.~C.~Hwa and X.~N.~Wang,
  River Edge, USA: World Scientific (2004) 777 p
\bibitem{whitebrahms}
 I.~Arsene {\it et al.}  [BRAHMS Collaboration],
perspective
  Nucl.\ Phys.\ A {\bf 757}, 1 (2005)
\bibitem{whitephobos}
  B.~B.~Back {\it et al.},
  Nucl.\ Phys.\ A {\bf 757}, 28 (2005)
\bibitem{whitestar}
  J.~Adams {\it et al.}  [STAR Collaboration],
quark  gluon
evidence from
  Nucl.\ Phys.\ A {\bf 757}, 102 (2005)
\bibitem{whitephenix}
 K.~Adcox {\it et al.}  [PHENIX Collaboration],
nucleus
collaboration,''
  Nucl.\ Phys.\ A {\bf 757}, 184 (2005)


\bibitem{alice}
  K.~Aamodt {\it et al.}  [ALICE Collaboration],
  Phys.\ Rev.\ Lett.\  {\bf 105}, 252302 (2010)
  [arXiv:1011.3914 [nucl-ex]].

\bibitem{cms}
  S.~Chatrchyan {\it et al.}  [CMS Collaboration],
  Phys.\ Rev.\ C {\bf 87}, 014902 (2013)
  [arXiv:1204.1409 [nucl-ex]].

\bibitem{atlas}
  G.~Aad {\it et al.}  [ATLAS Collaboration],
  Phys.\ Rev.\ C {\bf 86}, 014907 (2012)
  [arXiv:1203.3087 [hep-ex]].

\bibitem{Poskanzer:1998yz}
  A.~M.~Poskanzer and S.~A.~Voloshin,
  Phys.\ Rev.\ C {\bf 58}, 1671 (1998)
  [nucl-ex/9805001].

\bibitem{Heinz:2013th}
  U.~Heinz and R.~Snellings,
  Ann.\ Rev.\ Nucl.\ Part.\ Sci.\  {\bf 63}, 123 (2013)
  [arXiv:1301.2826 [nucl-th]].

\bibitem{littlebang}
  R.~Snellings,
  J.\ Phys.\ Conf.\ Ser.\  {\bf 381}, 012019 (2012).

%
\bibitem{Heinz:2013wva}
  U.~W.~Heinz,
  J.\ Phys.\ Conf.\ Ser.\  {\bf 455}, 012044 (2013)
  [arXiv:1304.3634 [nucl-th]].

\bibitem{gtscaling}
  G.~Torrieri,
  Phys.\ Rev.\ C {\bf 89}, 024908 (2014)
  [arXiv:1310.3529 [nucl-th]].

\bibitem{pihydro}
G. Barenblatt
``Scaling, self-similarity and intermediate asymptotics'', Cambridge texts in applied mathematics (2009)

\bibitem{bjorken}
  J.~D.~Bjorken,
  Phys.\ Rev.\ D {\bf 27}, 140 (1983).



\bibitem{Gubser:2010ze}
  S.~S.~Gubser,
  Phys.\ Rev.\ D {\bf 82}, 085027 (2010)
  [arXiv:1006.0006 [hep-th]].

\bibitem{Gubser:2010ui}
  S.~S.~Gubser and A.~Yarom,
  Nucl.\ Phys.\ B {\bf 846}, 469 (2011)
  [arXiv:1012.1314 [hep-th]].




\bibitem{Csanad:2005gv}
  M.~Csanad, T.~Csorgo, A.~Ster, B.~Lorstad, N.~N.~Ajitanand, J.~M.~Alexander, P.~Chung and W.~G.~Holzmann {\it et al.},
  Eur.\ Phys.\ J.\ A {\bf 38}, 363 (2008)
  [nucl-th/0512078].

\bibitem{cooperfrye}
  F.~Cooper and G.~Frye,
  Phys.\ Rev.\ D {\bf 10}, 186 (1974).

\bibitem{teaney}
  D.~Teaney,
  Phys.\ Rev.\ C {\bf 68}, 034913 (2003)
  [nucl-th/0301099].






\bibitem{Hatta:2014upa}
  Y.~Hatta and B.~-W.~Xiao,
  arXiv:1405.1984 [nucl-th].

 \bibitem{Molnar:2011kx}
  D.~Molnar,
  J.\ Phys.\ G {\bf 38}, 124173 (2011)
  [arXiv:1107.5860 [nucl-th]].


\bibitem{Dusling:2009df}
  K.~Dusling, G.~D.~Moore and D.~Teaney,
  Phys.\ Rev.\ C {\bf 81}, 034907 (2010)
  [arXiv:0909.0754 [nucl-th]].

\bibitem{Bhalerao:2013pza}
  R.~S.~Bhalerao, A.~Jaiswal, S.~Pal and V.~Sreekanth,
  Phys.\ Rev.\ C {\bf 89}, 054903 (2014)
  [arXiv:1312.1864 [nucl-th]].

\bibitem{Marrochio:2013wla}
  H.~Marrochio, J.~Noronha, G.~S.~Denicol, M.~Luzum, S.~Jeon and C.~Gale,
  arXiv:1307.6130 [nucl-th].


\bibitem{Bhalerao:2005mm}
  R.~S.~Bhalerao, J.~-P.~Blaizot, N.~Borghini and J.~-Y.~Ollitrault,
  Phys.\ Lett.\ B {\bf 627}, 49 (2005)
  [nucl-th/0508009].


\bibitem{Kolb:1999it}
  P.~F.~Kolb, J.~Sollfrank and U.~W.~Heinz,
  Phys.\ Lett.\ B {\bf 459}, 667 (1999)
  [nucl-th/9906003].



\bibitem{Staig:2011wj}
  P.~Staig and E.~Shuryak,
  Phys.\ Rev.\ C {\bf 84}, 044912 (2011)
  [arXiv:1105.0676 [nucl-th]].


\bibitem{Lacey:2013is}
  R.~A.~Lacey, Y.~Gu, X.~Gong, D.~Reynolds, N.~N.~Ajitanand, J.~M.~Alexander, A.~Mwai and A.~Taranenko,
  arXiv:1301.0165.

\bibitem{Gorda:2014msa}
  T.~Gorda and P.~Romatschke,
  arXiv:1406.6405 [nucl-th].

   \bibitem{Teaney:2012ke}
  D.~Teaney and L.~Yan,
  Phys.\ Rev.\ C {\bf 86}, 044908 (2012)
  [arXiv:1206.1905 [nucl-th]].

\bibitem{Floerchinger:2013rya}
  S.~Floerchinger and U.~A.~Wiedemann,
  Phys.\ Lett.\ B {\bf 728}, 407 (2014)
  [arXiv:1307.3453 [hep-ph]].



\bibitem{Borghini:2005kd}
  N.~Borghini and J.~-Y.~Ollitrault,
  Phys.\ Lett.\ B {\bf 642}, 227 (2006)
  [nucl-th/0506045].


\bibitem{Monnai:2009ad}
  A.~Monnai and T.~Hirano,
  Phys.\ Rev.\ C {\bf 80}, 054906 (2009)
  [arXiv:0903.4436 [nucl-th]].

\bibitem{Dusling:2011fd}
  K.~Dusling and T.~Sch\"afer,
  Phys.\ Rev.\ C {\bf 85}, 044909 (2012)
  [arXiv:1109.5181 [hep-ph]].

\bibitem{Noronha-Hostler:2013gga}
  J.~Noronha-Hostler, G.~S.~Denicol, J.~Noronha, R.~P.~G.~Andrade and F.~Grassi,
  Phys.\ Rev.\ C {\bf 88}, 044916 (2013)
  [arXiv:1305.1981 [nucl-th]].

\bibitem{Noronha-Hostler:2014dqa}
  J.~Noronha-Hostler, J.~Noronha and F.~Grassi,
  arXiv:1406.3333 [nucl-th].

\bibitem{Denicol:2014vaa}
  G.~S.~Denicol, S.~Jeon and C.~Gale,
  arXiv:1403.0962 [nucl-th].


\bibitem{alicemult}
  K.~Aamodt {\it et al.}  [ALICE Collaboration],
  Phys.\ Rev.\ Lett.\  {\bf 105}, 252301 (2010)
  [arXiv:1011.3916 [nucl-ex]].

\bibitem{alicept}
  B.~B.~Abelev {\it et al.}  [ALICE Collaboration],
  Phys.\ Lett.\ B {\bf 727}, 371 (2013)
  [arXiv:1307.1094 [nucl-ex]].



\bibitem{scale}
 S.~Chatrchyan {\it et al.}  [CMS Collaboration],
  Phys.\ Lett.\ B {\bf 724}, 213 (2013)
  [arXiv:1305.0609 [nucl-ex]].



\bibitem{paridge1}
  A.~Adare {\it et al.}  [PHENIX Collaboration],
  Phys.\ Rev.\ Lett.\  {\bf 111}, 212301 (2013)


\bibitem{hsong}
  H.~Song and U.~W.~Heinz,
  Phys.\ Rev.\  C {\bf 78}, 024902 (2008)
  [arXiv:0805.1756 [nucl-th]].

\bibitem{scanv2paper}
  L.~Adamczyk {\it et al.}  [STAR Collaboration],
  Phys.\ Rev.\ C {\bf 86}, 054908 (2012)
  [arXiv:1206.5528 [nucl-ex]].

\bibitem{kestin}
  G.~Kestin and U.~W.~Heinz,
  Eur.\ Phys.\ J.\ C {\bf 61}, 545 (2009)
  [arXiv:0806.4539 [nucl-th]].


\bibitem{alicevn}
  K.~Aamodt {\it et al.}  [ALICE Collaboration],
  Phys.\ Rev.\ Lett.\  {\bf 107}, 032301 (2011)
  [arXiv:1105.3865 [nucl-ex]].

\bibitem{aaridgeatlas}
  G.~Aad {\it et al.}  [ATLAS Collaboration],
  Phys.\ Rev.\ C {\bf 86}, 014907 (2012)
  [arXiv:1203.3087 [hep-ex]].


\bibitem{heinzvn}
  U.~Heinz, Z.~Qiu and C.~Shen,
  Phys.\ Rev.\ C {\bf 87}, no. 3, 034913 (2013)
  [arXiv:1302.3535 [nucl-th]].




\bibitem{highptscaling}
  A.~Adare {\it et al.}  [PHENIX Collaboration],
  Phys.\ Rev.\ Lett.\  {\bf 98}, 162301 (2007)
  [nucl-ex/0608033].

\bibitem{paridge2}
  G.~Aad {\it et al.}  [ATLAS Collaboration],
  Phys.\ Rev.\ Lett.\  {\bf 110}, 182302 (2013)

\bibitem{keanestock}
  S.~Wang, Y.~Z.~Jiang, Y.~M.~Liu, D.~Keane, D.~Beavis, S.~Y.~Chu, S.~Y.~Fung and M.~Vient {\it et al.},
  Phys.\ Rev.\ C {\bf 44}, 1091 (1991).

\bibitem{factor}
  F.~G.~Gardim, F.~Grassi, M.~Luzum and J.~-Y.~Ollitrault,
  Phys.\ Rev.\ C {\bf 87}, no. 3, 031901 (2013)
  [arXiv:1211.0989 [nucl-th]].

\bibitem{raju}
  B.~Schenke, P.~Tribedy and R.~Venugopalan,
  Phys.\ Rev.\ Lett.\  {\bf 108}, 252301 (2012)
  [arXiv:1202.6646 [nucl-th]].

\bibitem{bron}
  J.~-P.~Blaizot, W.~Broniowski and J.~-Y.~Ollitrault,
  arXiv:1405.3274 [nucl-th].

\bibitem{Andrade:2014swa}
  R.~P.~G.~Andrade, J.~Noronha and G.~S.~Denicol,
  arXiv:1403.1789 [nucl-th].

\bibitem{kolb}
  P.~F.~Kolb and U.~W.~Heinz,
  In *Hwa, R.C. (ed.) et al.: Quark gluon plasma* 634-714
  [nucl-th/0305084].

\bibitem{pratt}
  J.~Novak, K.~Novak, S.~Pratt, C.~Coleman-Smith and R.~Wolpert,
  arXiv:1303.5769 [nucl-th].

\bibitem{heinz}
  C.~Shen, S.~A.~Bass, T.~Hirano, P.~Huovinen, Z.~Qiu, H.~Song and U.~Heinz,
  J.\ Phys.\ G {\bf 38}, 124045 (2011)
  [arXiv:1106.6350 [nucl-th]].

\bibitem{bass}
  G.~-Y.~Qin, H.~Petersen, S.~A.~Bass and B.~Muller,
  Phys.\ Rev.\ C {\bf 82}, 064903 (2010)
  [arXiv:1009.1847 [nucl-th]].























\end{thebibliography}
\end{document}